# Aging Study of RPC's for the SiD Hcal and Muon System

Changguo Lu, Kirk McDonald, and A. J. S. Smith

*Princeton University, Jadwin Hall, Washington Road, Princeton, NJ 08544, USA*

and

Jiawen Zhang[2] [*]

*Institute of High Energy Physics, 19 Yu Quang Rd, Beijing 100041, P. R. China*

Preliminary test results on microscope investigation and BESIII-type RPC aging performance have revealed interesting aging phenomena that had not been seen before in Linseed oil coated Italian-type RPC. We report here on the aging performance of BESIII-type and its variant RPC, and on microscopic surface characterization of BESIII-type Bakelite electrodes.

## 1 Introduction

RPCs built from the new type of Bakelite developed by the BESIII Muon group of IHEP (Beijing) and Gaonengkedi, Inc. (Beijing) for use in the BES III and Daya Bay Muon Systems [1] have achieved acceptable dark noise rates without Linseed oil coating, but aging effects have not been thoroughly studied - there is no published report available on this topic. A preliminary study of the Daya Bay Muon System RPCs has indicated a significant aging effect [2], which might be adequate for underground neutrino experiments, but must be understood and mitigated prior to use of this technology for SiD and other future accelerator experiments.

## 2 Aging test of BESIII-type RPC

We have used 5 BESIII-type 50cm×50cm RPCs in the test. These five RPCs were made of the same Bakelite as BESIII and Daya Bay muon detectors.

### 2.1 Test set-up

Figure 1 shows a photo of our set-up for RPC aging tests and cosmic ray performance studies.

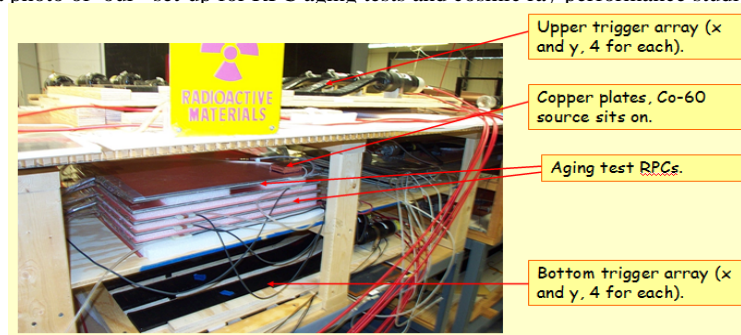

**Fig. 1. RPC aging test set-up.**

[*] This work is supported by the US DoE Program of University Linear Collider Detector R&D, Project 6.19.

The gas inter-connection among five RPCs is shown in Fig. 2, and figure 3 shows the scintillation counter array used for cosmic ray trigger. It sandwiches the stack of test RPCs, and divides the test RPCs into 16 regions (#1 to #16 from lower-right to upper-left) in a 4×4 array.

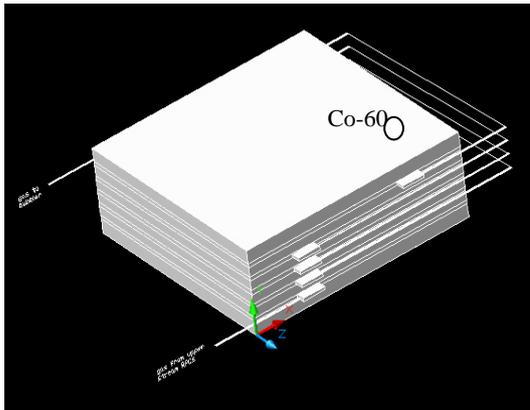

**Fig. 2. Gas inter-connection among five RPCs and the source position.**

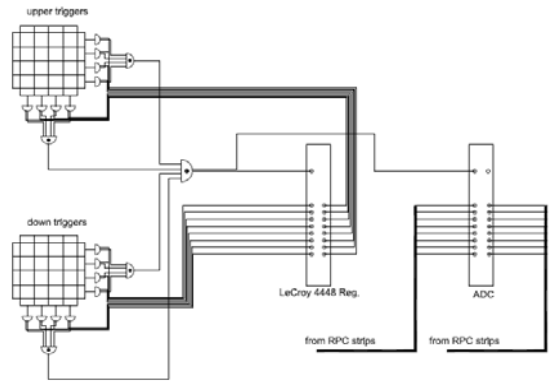

**Fig. 3. Scintillation telescope array.**

## 2.2    Singles rate

A 0.1mCi Co-60 source is placed on top of the RPCs through an 18mm thick stack of copper plates (region #1, see Fig. 2). The singles rates with and without the source are quite different among 5 RPCs. RPC #4 is the noisiest, ~6kHz for strip #2. RPC #8 is the quietest, ~0.8kHz for strip #2. See figure 4 for details.

Because of the geometry and the intrinsic noise rate difference the aging dose rate is different among these RPCs, with their ratio is (#8): (#6, #7): (#4) ~ 1: 2.5: 7.5. The equivalent aging dose is based on the following assumptions: (1) The background noise rate for a running RPC is 0.4Hz/cm$^2$; and (2) A simple Monte Carlo calculation shows 40% of the total rate detected by 8 6cm-width strips over a 50cm×50cm RPC by a Co-60 source that is concentrated in a 10cm radius circle in region #1  [2].

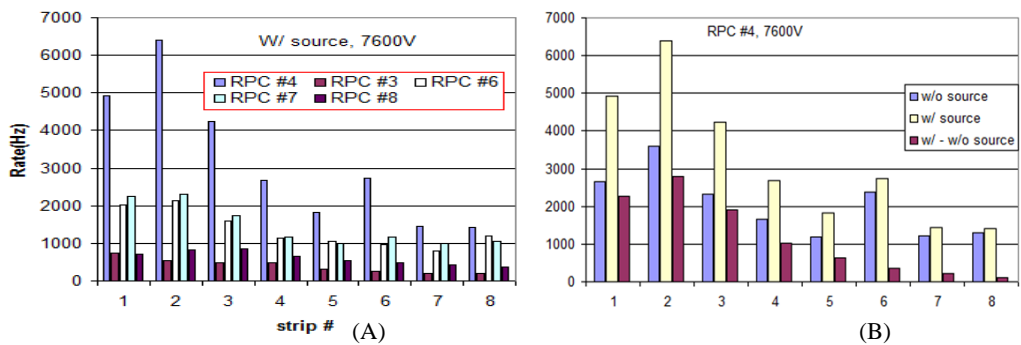

**Fig. 4. Singles rate of the test RPCs. (A) Rate for all five RPCs with source; (B) Rate for RPC#4 with and without source.**

## 2.3    Efficiency test

For the aging test the source is placed on top of the RPC stack. We measured the initial efficiency for the test RPCs, monitored their dark current in the course of aging test. By comparing the efficiency degradation with time the aging effect can be clearly revealed. Figure 5 shows the efficiency before and after 23 days aging. The equivalent aging dose for RPC#4 equals to 7.6 years. By that time serious aging has already shown up. The other three RPCs have much less degradation because their equivalent aging dose is smaller.

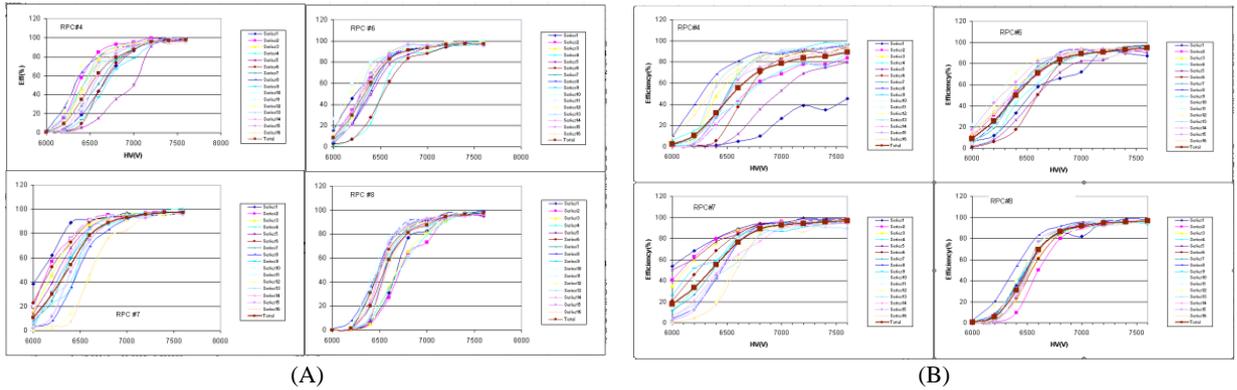

(A)                                                   (B)

**Fig. 5(A) Initial efficiency; (B) After 23 days efficiency.**

## 2.4    Linseed oil coated BESIII-type RPC

From our previous tests we have observed that the Linseed oil coating can protect the Bakelite surface from HF attack [3]. Experience with Italian-type RPC shows better aging performance that confirms our observation. To obtain direct confirmation we coated the inner surface of a BESIII-type RPC with diluted Linseed oil (35% Linseed oil + 65% of n-pentane), the coating thickness is ~ 7μm.

The initial performance of this chamber is extremely good. Its singles rate with and without source is shown in Fig. 6. The noise rate without source is very close to cosmic ray background. The equivalent aging dose ratio to the predefined background rate (0.4Hz/cm$^2$) due to the source is estimated as 40:1.

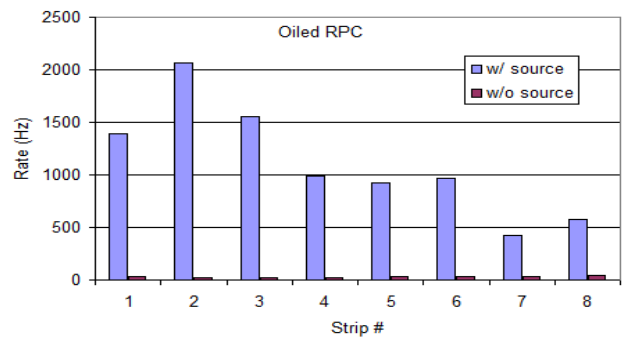

**Fig. 6. Intial singles rate of  Linseed oil coated RPC.**

In figure 7 we compare the aging performance for Linseed oil coated RPCs and other two RPCs: RPC#4 and #8. The number in the bright yellow box is the equivalent aging dose. Up to 11 years of the equivalent aging dose the oil coated RPC still has good efficiency curve, although the efficiency plateau's shoulder has shifted higher by ~ 400V. For other two regular RPCs without oil coating (7.6 and 5.7 years, respectively) the efficiency plateaus already degraded profoundly in spite of significantly less aging dose.

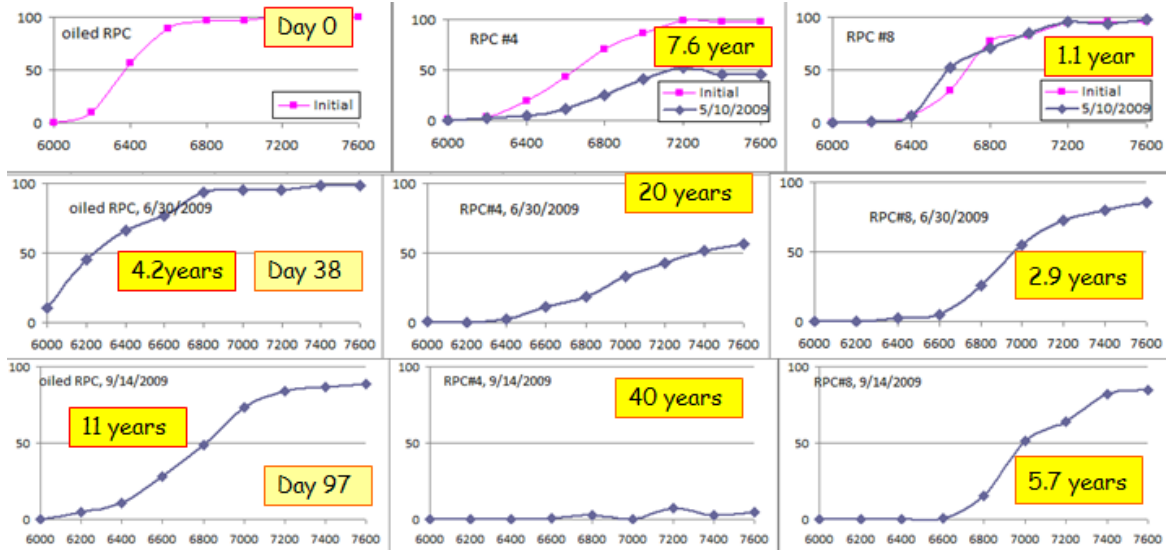

**Fig. 7. Comparison between oil coated BESIII RPC and regular BESIII RPC.**

A month after the Day 97 of the efficiency test, when we tested the efficiency again we found drastic efficiency drop for the radiated region of the oil coated RPC. An "autopsy" revealed that the Linseed oil coating had not been cured completely, causing several big oil stalagmites to form in the gap of the radiated region, similarly to the serious failure of the first generation of BaBar RPCs. [4]. We believe if we improve the curing process, this problem can be avoided as demonstrated by second generation of BaBar RPC's [5, 6]. The advantage of the Linseed oil coating is clear: lower noise rates and better aging performance.

### 2.5    Oil embedded Bakelite RPC

Based on our test results our Chinese collaborator, Xianghu Construction Material, Inc. has developed a new variant of Bakelite with embedded Linseed oil. Five RPCs with this Bakelite have been constructed and put into aging test for more than 75 days, an aging dose equivalent to more than 9 years. The most recent test results are summarized in figure 7.

To simplify the plateau plots instead of all 16 regions we only draw three regions for each RPC: region #1, #2 and #5, which are the direct radiated regions and their immediate neighboring regions. The equivalent aging doses are shown in the right yellow text boxes. Surprisingly RPC N5 efficiency is even getting better with time, and the other four RPCs are also showing no aging degradation. At the 74-th day the equivalent aging dose reaches 8.2 years. On other hand for the regular BESIII-type RPC, at equivalent aging dose of 7.6 years serious aging has already taken place, as shown in figure 5.

# 3    Microscope study on various aged Bakelite electrodes

## 3.1    Microscope and pin-probe station

A Nikon LV100D microscope has been used for this study to search for the cause of the aging. We have set up a microscope/pin-probe station aimed the resistance measurement that has revealed tiny sparking marks on the electrode surface, as shown in figure 8.

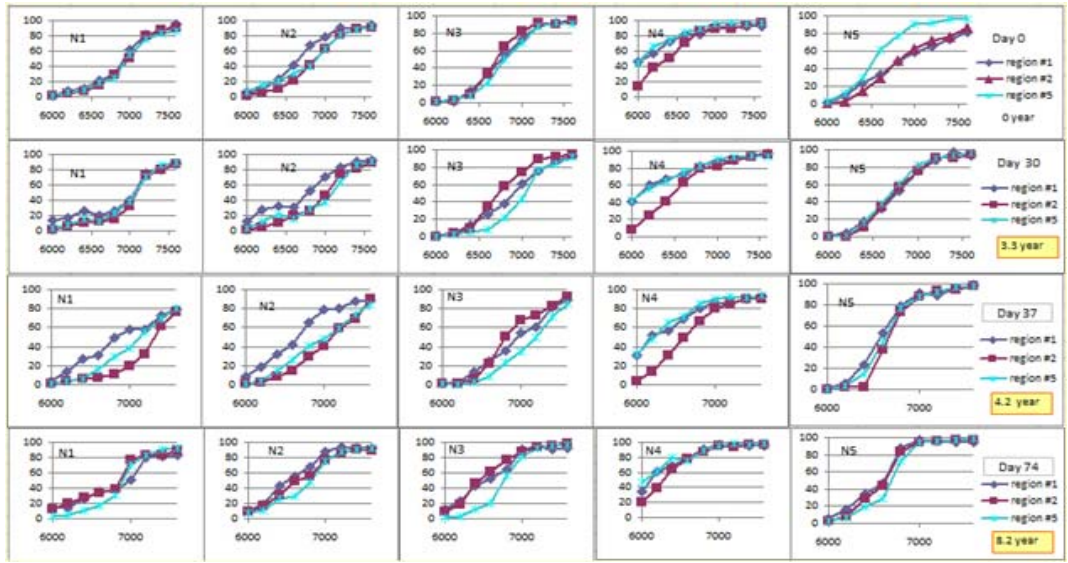

**Fig. 7. Summary of the Linseed oil embedded Bakelite RPC aging test.**

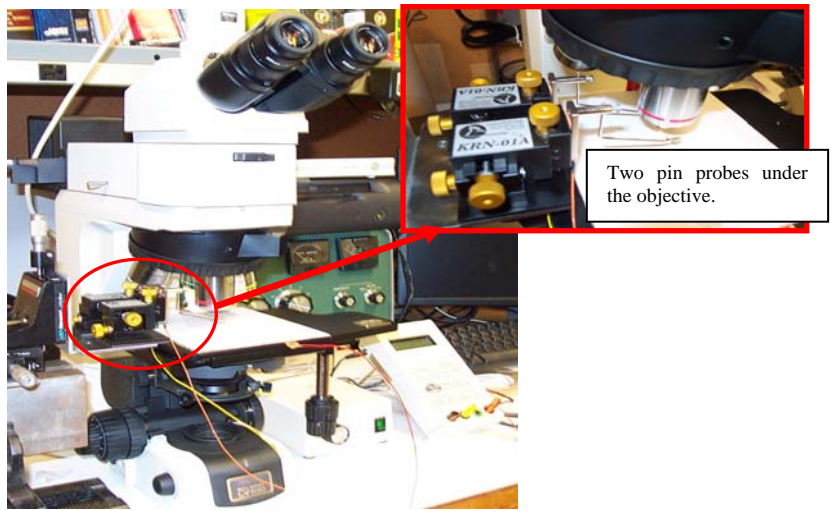

Two pin probes under the objective.

**Fig. 8. Nikon microscope and two pin-probes with their 3D micrometer stages (the insert shows two pin-probes).**

## 3.2 Aged RPC electrode surface images

After opening the aged RPCs we found a lot of sparking marks all over the surface for both anode and cathode. Figure 9 shows some typical images. The distribution density of the marks varies; and shows a correlation between the degree of efficiency drop and the density of the sparking marks. RPC #4 has suffered the highest aging dose, and the highest sparking density. RPC #8 has the least aging dose and lowest aging mark density. Under the microscope the sparking marks show varied appearances that may be related to the degree of the electric energy released in the sparks.

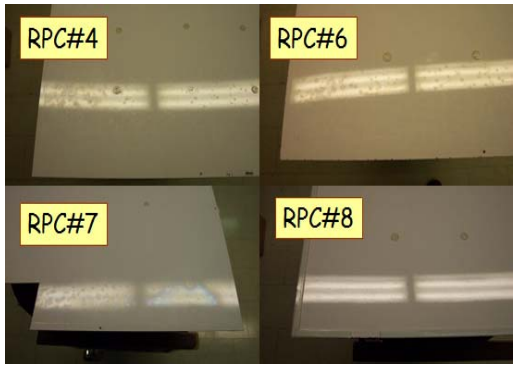

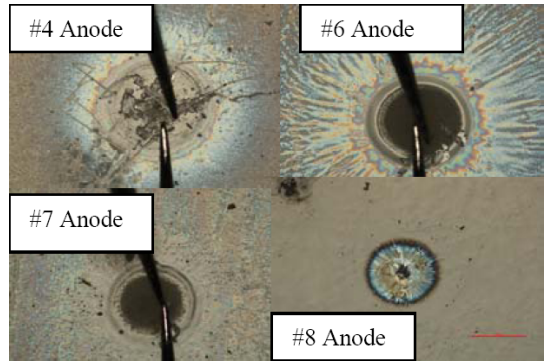

**Fig. 9. Sparking marks on the aged RPC electrode surface can be seen in the ceiling lamp's images.**

**Fig. 10. Sparking mark images from four RPCs: #4, #6, #7 and #8. The 0.5mm red microscope scale is shown in #8 image.**

In figure 10 we show images of typical sparking mark images for different RPCs. The sharp pins shown in the picture are the pin probes. Scratching the sparking marks with the probe can produce clear scratching traces. By contrast there is very little sparking evidence found on the Linseed oil coated RPC electrodes. In Fig. 11 we show images of spark-like spots, but their different appearance may indicate that the Linseed oil coating plays a big role in suppressing large sparking.

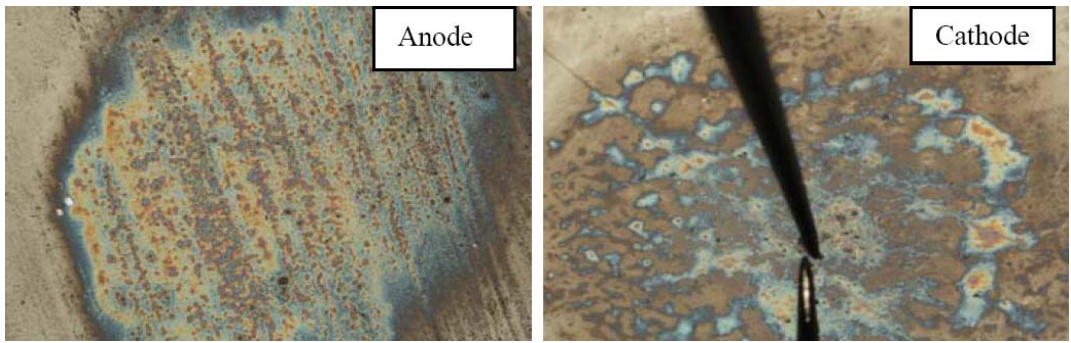

**Fig. 1. Spark-like marks observed on Linseed oil coated surface.**

### 3.3    Volume resistance study for the Bakelite electrode of the aging test RPCs

Using the microscope/pin probe station we can measure the volume resistance for any interested region on the electrode surface. We can move the probes to the sparking region, place the pins right on the interested region, and lower the probes to touch the surface as shown in figure 10 and 11. Two pin probes are connected together and form one input to the high resistance meter. The other input of the resistance meter is connected to a copper tape that sticks to the graphite coating on the back side of the test sample. Two categories of the interested region are selected: sparking and non-sparking areas. The test chambers have their original Bakelite volume resistance measured and labeled. We summarize all of the test results in figure 12. There is some small resistance difference between sparking and non-sparking areas, the former is somewhat smaller than the later, but it is not dramatic and whether or not it is related to the efficiency degradation is still unknown.

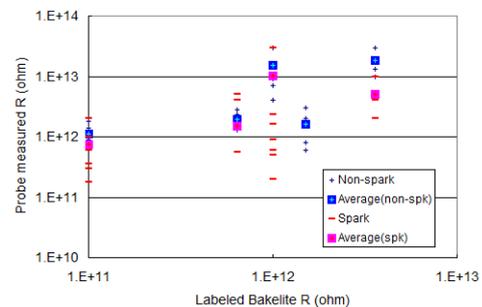

**Fig. 12. Pin probe measured volume resistance for various aged electrodes.**

# 4    BaBar RPC study[a]

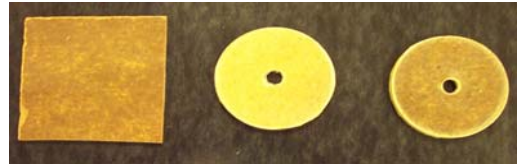

**Fig. 13. BaBar RPC sample discs.**

The new generation RPCs of the BaBar endcap muon detector system had been operating at PEPII for five and half years since November 2002 [5, 6]. The circular ring region surrounding the beam pipe endured the highest particle flux and showed serious efficiency loss. It was therefore of special interest to study these samples and determine the cause of the aging. Figure 13 shows three samples of BaBar Bakelite: at the left is new Bakelite, in the middle is the sample from the high particle flux region, and at the right is the sample from the lower flux region.

The sample disc from the high flux central region shows a highly discolored surface with an "orange peel" looking texture. The disc from the lower flux region is less discolored, its Linseed oil coating film looks smooth, and a cotton tip with ethanol easily removes the oil film, revealing a shiny Bakelite surface. It seems that the high dose of radiation and resulted erosion from high concentration of HF caused the dye in the surface layer paper to discolor; and also caused the "orange peel" look on the surface. Figure 14 shows a comparison between high dose and low dose BaBar samples.

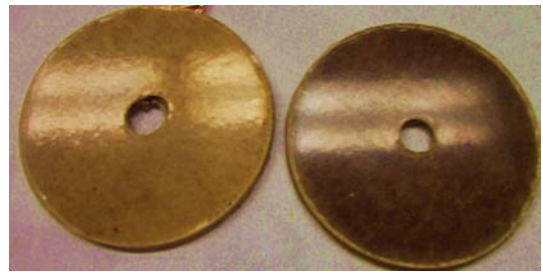

**Fig. 14. The left disc is high dose sample, it shows discolored "orange peel" like surface, the right disc is low dose sample, its surface looks reasonably smooth and darker.**

## 4.1    Surface resistance change

A Trek Model 152-1-CE high resistance meter together with Model 152-2p two-point resistance probe is used to measure the surface resistance for these BaBar sample discs. The probe's two contacts are 3.2mm in diameter, 6.4mm in separation. The test results are summarized in Table 1.

These discs come from same RPC so we can assume their initial surface resistivity was approximately the same. Table 1 shows the surface resistance measured for high dose samples is less than from low dose samples, the ratio is ~ 1:5.

| Test sample | | | Resistance ($\Omega$) | Average R ($\Omega$) |
|---|---|---|---|---|
| BT26-6 | High dose | Highly discolored | $6.6 \times 10^{10}\,\Omega$ | $6.0 \times 10^{10}\,\Omega$ |
| BT23-1 | | | $5.3 \times 10^{10}\,\Omega$ | |
| BT14-1 | Low dose | Less discolored | $2.5 \times 10^{11}\,\Omega$ | $2.8 \times 10^{11}\,\Omega$ |
| Bt10-1 | | | $3.0 \times 10^{11}\,\Omega$ | |

**Table 1. Surface resistance of the BaBar sample discs.**

[a] The authors would like to thank Henry Band for providing the BaBar aged RPC samples that made this study possible.

## 4.2 Surface microscope study

Compared to the BESIII RPC we notice that sparking marks are very rare on our BaBar RPC samples. We did find one and its microscope image is shown in figure 15, quite different looking from BESIII-type RPC.

The different appearance of the sparking marks for Linseed oil coated BaBar RPCs and oil-free BESIII-type RPCs might be related to their different surface material. The Linseed oil coating, when it is completely cured, has a long carbon chain, as shown in figure 16(A) [7], and the melamine structure, which is the surface material of BESIII Bakelite, is shown in figure 16(B) [8].

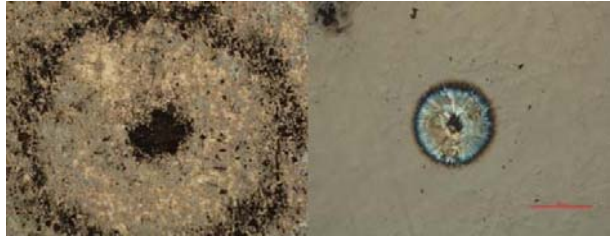

**Fig. 15. Sparking mark, (A) BaBar sample; (B) BESIII-type aging test chamber #8 sample. The red microscope scale (0.5mm) is shown in (B), it should be same for (A).**

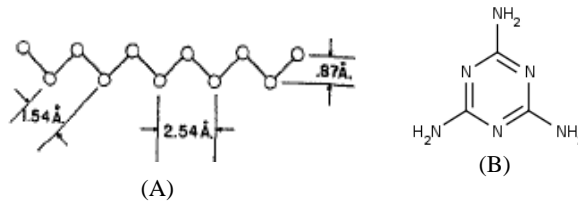

**Fig. 16. Chemical structure of Bakelite surface material. (A) cured Liseed oil film; (B) Melamine resin.**